\documentstyle [12pt]{article}
\newcommand{\pbarp}{\overline{p}p}
\newcommand{\qqbar}{Q\overline{Q}}
\newcommand{\KKbar}{K\overline{K}}

\newcommand{\nnbar}{\overline{n}n}
\newcommand{\uubar}{u\overline{u}}
\newcommand{\ddbar}{d\overline{d}}
\newcommand{\ssbar}{s\overline{s}}

\begin{document}
\title{\small \rm \begin{flushright} \small{hep-ph/9505219}\\
\small{RAL-95-036}\\
3 May 1995 \end{flushright} \vspace{2cm}
\LARGE {\bf Evidence for a Scalar Glueball}
 %#?
\vspace{0.8cm} }
\author{Claude Amsler\thanks{E-mail : amsler@cernvm.cern.ch}\\
{\small \em Physik-Institut, Universit\"at Z\"urich, CH-8057 Z\"urich,
Switzerland} \\ \\
 Frank E. Close\thanks{E-mail : fec@v2.rl.ac.uk}\thanks{Work supported
in
part by the European Community Human Capital Mobility program
``Eurodafne",
Contract CHRX-CT92-0026} \\
{\small \em Particle Theory, Rutherford-Appleton Laboratory, Chilton,
Didcot OX11 0QX, UK} \\  \\}
\date{May1995 \vspace{1.5cm}}

\begin{center}
\maketitle

\begin{abstract}
We show that the newly discovered scalar meson $f_0(1500)$ at LEAR
has properties compatible with the lightest scalar glueball predicted by
lattice QCD and incompatible with a $\qqbar$ state. We suggest that
decays of glueballs are into pairs of
glueballs (including $\eta, \eta'$ or $(\pi\pi)_S$) or by mixing with
nearby $Q\bar{Q}$ states. The partial widths
of $f_0(1500)$ are in accord with this hypothesis, tests of which
include characteristic radiative decays to $\gamma\phi,\gamma\omega,
\gamma\rho$ and the prediction of a further scalar state, $f_0'(1500 -
1800)$ which couples strongly to $K\bar{K}$, $\eta\eta$
and $\eta\eta'$.
\end{abstract}

\vspace*{1mm}
Submitted to Physics Letters

\end{center}

%%%%%%%%%%%%%%%%%%%%%%%%%%%%%%%%%%%%%%%%%%%%%%%%%
\newpage

Glueballs are a major missing component of the standard model.
For two decades, experimental searches
for bound states of gluons have produced only controversial signals.
Whereas the gluonic degrees of freedom expressed in $L_{QCD}$
have been established beyond doubt in high momentum
data, their dynamics in the strongly interacting limit epitomised by
hadron spectroscopy remain obscure.

In this letter we show that if intuition based on lattice QCD is a reliable
guide, the emerging properties of the scalar meson
$f_0(1500)$ discovered at LEAR are consistent with those of a
glueball mixed with nearby members of the quark model scalar nonet.
This hypothesis may be tested in forthcoming experiments.

There are now clear signals for the $f_0(1500)$ in
a range of production processes that are
traditionally believed to favour glueballs, namely \cite{closerev}
\begin{enumerate}
\item
Radiative $J/\psi$ decay: $J/\psi  \rightarrow \gamma+G$ \cite{bugg}
\item
Collisions in the central region away from quark beams and target:
$pp \rightarrow p_f(G)p_s$ \cite{Kirk,Gentral}.
\item
Proton-antiproton annihilation where the destruction of quarks creates
opportunity for gluons to be manifested
\cite{Anis}-\cite{Hasan1}. The $f_0(1500)$ was observed to
decay into $\pi^0\pi^0$, $\eta\eta$ and $\eta\eta'$
in  $\pbarp \rightarrow $ $3\pi^0$ \cite{Anis,Amsler3pi0},
$\pi^0\eta\eta$  \cite{Augustin,Amsleretaeta}
and $\pi^0\eta\eta'$ \cite{Enhan}.
\end{enumerate}

\noindent By contrast, there are no significant sightings of $f_0(1500)$
in processes where glueballs are not expected to be
enhanced.

This prima facie evidence for a scalar glueball is particularly interesting
now that studies of lattice-QCD appear to be coming to a concensus
that the lightest glueball is indeed a scalar in this region of mass
\cite{ukqcd,weing}. In these circumstances it is natural to
speculate that the $f_0(1500)$ and the ``primitive" (i.e. quenched
approximation) scalar glueball of lattice-QCD are intimately related.
The purpose of this letter is to evaluate and to propose further
tests of this hypothesis. A detailed discussion may be found in
 ref. \cite{cafe}.

Lattice QCD is not yet able to make detailed quantitative statements
about the partial decay widths of glueballs against which we could
evaluate those of $f_0(1500)$. However, qualitative features may be
abstracted from flux-tube models which are probably the nearest we
can presently get to simulating the strong coupling lattice
formulation of QCD. In this simulation \cite{paton85} mesons
consist of a quark and antiquark connected by a tube of coloured
flux whereas glueballs G$_0$ consist of a loop of flux.
These eigenstates are perturbed by two
types of interaction \cite{kokoski87}

\begin{enumerate}
\item
$V_1$ which creates a $Q$ and a $\bar{Q}$ at neighbouring lattice
sites, together with an elementary flux-tube connecting them. When
$V_1$ acts on conventional $Q\bar{Q}$ mesons it causes their decays
(fig. \ref{Rivsx}a), and is thereby known to be
significant. Fig. \ref{Rivsx}b shows that when $V_1$ acts on a
{\bf glueball} G$_0$, in leading order
 it causes {\bf glueball -$Q\bar{Q}$ mixing}.
\item
$V_2$ which creates or destroys a unit of flux around any plaquette,
where a plaquette is an elementary square with links on its edges: in
leading order this causes glueballs to decay to {\bf pairs of glueballs}
(fig. \ref{Rivsx}c).
\end{enumerate}

\noindent Note that in this approach
$G_0$ does not decay to $\pi\pi$ or $K\bar{K}$
in leading order since gluons are isoscalar;
 decays to $\eta$ and $\eta'$ are allowed to the
extent that these have nonzero overlap with glue.

The mixing with $Q\bar{Q}$ is likely to play a significant role in glueball
phenomenology if a quarkonium nonet of the same $J^{PC}$ is nearby.
As a result of extensive data from the Crystal Barrel collaboration
at LEAR it is now clear that there is a scalar nonet in the 1.5 GeV region,
as well as the $f_0(1500)$.

(i) The discovery of the $I$ = 1 state
 $a_0(1450)\rightarrow\eta\pi$ \cite{Spanier}
sets the natural scale of masses and widths for the other members of
the scalar nonet. In quark models such as ref. \cite{kokoski87,Godfrey} the
widths of the scalar $Q\bar{Q}$ are qualitatively ordered as
$\Gamma(\nnbar) > \Gamma(\ssbar) > \Gamma(a_0) \geq
\Gamma(K^*)$.
Empirically $\Gamma(a_0)$ = 270 $\pm$ 40 MeV,
$\Gamma(K^*_0)$ = 287
$\pm$ 23 MeV which are consistent with quark model expectations and
lead one to expect for their partners
$\Gamma(\nnbar)\sim$ 400-700  MeV and $\Gamma(\ssbar)
\sim$ 300-500 MeV \cite{cafe,kokoski87,Godfrey,Barnes}.

(ii) The Crystal Barrel data show clear signals for an
independent scalar meson $f_0(1370)$ in $\pi\pi$ and $\eta\eta$
whose mass is consistent with it being the $n\bar{n}$ partner of the
$a_0(1450)$ and whose ratio of partial widths to $\pi\pi$ and $\eta\eta$
also is consistent with it being the
$n\bar{n} \equiv (\uubar +\ddbar)/\sqrt{2}$ state
of a nonet \cite{Anis,Amsler3pi0,Amsleretaeta,Spanier}.
The total width of $f_0(1370)$ is not yet well determined,
200-700 MeV being possible \cite{Anis,Amsleretaeta,Amsler3pi0}
depending on the theoretical model used in the analysis and in accord
with the $n\bar{n}$ hypothesis. The $\gamma\gamma$ width in this
region is also consistent with it containing the $n\bar{n}$ state
\cite{MP,BCL}. By contrast, the $f_0(1500)$ width  is 116 $\pm$ 17
MeV \cite{Amsler3pi0,Enhan,Amsleretaeta}
and is clearly out of line with the scalar nonet,
being even smaller than the $K^*$ and $a_0$ widths.

The properties of the $f_0(1500) - f_0(1370)$ system are
incompatible with them both belonging to a $Q\bar{Q}$ nonet.
The $f_0(1370)$ appears to be dominantly
$n\bar{n}$  and on mass grounds we expect that
the $s\bar{s}$ would lie some 200-300 MeV higher than this and the
$a_0(1450)$. The strong
coupling of $f_0(1500)$ to pions implies that it is not primarily an
$s\bar{s}$ state; the decoupling from $K\bar{K}$ further distances it
from the nonet. Specifically \cite{cafe} after correcting for phase space
and minor effects of form factors, the Crystal Barrel data imply for the
ratios of partial widths

\begin{equation}
R_1 \equiv
\frac{\gamma^2(f_0(1500)\rightarrow\eta\eta)}{\gamma^2(f_0(1
500)\rightarrow\pi\pi)} = 0.27\pm 0.11,
\label{R1}
\end{equation}
\begin{equation}
R_2 \equiv
\frac{\gamma^2(f_0(1500)\rightarrow\eta\eta')}{\gamma^2(f_0(1
500)\rightarrow\pi\pi)} = 0.19 \pm 0.08,
\label{R2}
\end{equation}
while a bubble chamber experiment leads to the (95\% C.L.) upper
limit \cite{Gray}
\begin{equation}
R_3 \equiv \frac{\gamma^2(f_0(1500)\rightarrow
K\bar{K})}{\gamma^2(f_0(1
500)\rightarrow\pi\pi)} < 0.1.
\label{R3}
\end{equation}

In the $\qqbar$ hypothesis it is possible
to fit two ratios but not all three for the $f_0(1500)$. This is shown in
fig.\ref{Glas} \cite{Amslerglas}:
The roughly equal couplings to $\eta\eta$ and $\eta\eta'$
together with the dominance of $\pi\pi$
(eqn.  \ref{R1} and \ref{R2}) imply that $f_0(1500)$, if $\qqbar$,  is
nearly $\nnbar$.
However, for an $\nnbar$ state, $\gamma^2(\KKbar)$ = 1/3
$\gamma^2(\pi\pi)$, in contradiction with data (eqn. \ref{R3}). Furthermore,
the strong affinity of
$f_0(1370)$ for $\pi\pi$ and its branching ratios and widths
suggest that this state is strongly $n\bar{n}$ which creates further
problems for a $Q\bar{Q}$ interpretation of $f_0(1500)$.   This
remains true for any reasonable breaking of $SU(3)_f$ symmetry
\cite{cafe}. These results are stable against form factor choice
\cite{cafe,Barnes} and contrast the experience with other known nonets
\cite{dok95}.

However, the observed decay branching ratios and in
particular the suppression of $\KKbar$  for the $f_0(1500)$
are natural for a scalar glueball which is in the vicinity of the
$Q\bar{Q}$ nonet and mixes with the two nearby
$\qqbar$ isoscalars, one ($n\bar{n}$) with mass below 1.5 GeV and the
other ($s\bar{s}$) above. The $f_0(1370)$ is a natural candidate for the
$n\bar{n}$; this hypothesis requires that a (mainly) $\ssbar$ state lies
in the 1600 MeV region.

We demonstrate this more quantitatively by considering the effect of
$V_1$ on a primitive glueball $G_0$. For this first look we assume
flavour blindness at the fundamental level such that
$\langle s\bar{s} |V_1| G_0 \rangle \equiv
\langle d\bar{d} |V_1| G_0 \rangle$.  The quarkonium mixing into the
glueball state is then

\begin{eqnarray}
 N_G |G \rangle =  |G_0 \rangle + \xi \{\sqrt{2}|n\bar{n}\rangle  +
 \omega  |s\bar{s} \rangle \}
\equiv |G_0 \rangle + \sqrt{2}\xi |\qqbar\rangle \nonumber\\
\label{3states}
\end{eqnarray}
where $N_G$ is the normalisation $\equiv \sqrt{1+\xi^2(2+\omega^2)}$,
 $\xi$ is the dimensionless mixing parameter

\begin{equation}
\xi \equiv \frac{\langle d\bar{d}|V_1|G_0 \rangle }{E_{G_0}-
E_{n\bar{n}}},
\end{equation}
and
\begin{equation}
\omega \equiv \frac{E_{G_0}-E_{n\bar{n}}}{ E_{G_0}-E_{s\bar{s}}}
\label{omega}
\end{equation}
 is the ratio
of the mass gap of $G_0$ and the $n\bar{n}$ and $s\bar{s}$
intermediate states in old fashioned
perturbation theory (fig. \ref{3graphs}).

Only in the particular case $\omega
=1$, where $E_{n\bar{n}} \equiv E_{s\bar{s}}$,
 does the underlying flavour blindness naturally survive at hadron
level as $G_0$ mixes into the flavour singlet
\begin{equation}
|\qqbar \rangle \equiv |u\bar{u} +d\bar{d} +s\bar{s} \rangle /\sqrt{3}.
\end{equation}
However, when $\omega \neq 1$, as will tend to be the case when
$G_0$ is in the vicinity of a $Q\bar{Q}$ nonet with the same $J^{PC}$
quantum numbers (as in the $0^{++}$ case of interest here)
significant distortion from naive flavour singlet can arise: the mass
breaking, $\Delta m \equiv m_s - m_d$, which is usually regarded as a
small perturbation in hadron dynamics, is here magnified.

There are three data that suggest a self consistency with this hypothesis.
\begin{enumerate}
\item
The suppression of $K\bar{K}$ in the $f_0(1500)$ decays suggests a
destructive interference between $n\bar{n}$ and $s\bar{s}$ such that
$\omega \approx -1$. This arises naturally if the
primitive glueball mass is equidistant between those of $n\bar{n}$
and the primitive $s\bar{s}$ - a situation not inconsistent with lattice
QCD. As the mass of $G_0 \rightarrow m_{n\bar{n}}$ or $m_{s\bar{s}}$,
the $K\bar{K}$ remains suppressed though non-zero. Specifically from
eqn. \ref{3states} and $SU(3)_f$ we obtain for $G \equiv f_0(1500)$
\begin{equation}
\langle \KKbar|V_1|G\rangle = \frac{1+\omega }{2}
\langle \pi\pi |V_1|G\rangle
\end{equation}
per unit charge combination for $\KKbar$ and $\pi\pi$.
The upper limit $R_3$ (eqn. \ref{R3}) leads to the constraint

\begin{equation}
-0.27 < \frac{\langle K\bar{K}|V_1|G
\rangle}{\langle \pi\pi | V_1|G \rangle} < 0.27
\end{equation}
which then gives the range

\begin{equation}
-1.5 < \omega  < -0.5.
\label{range}
\end{equation}
Thus eventual
quantification of the $K\bar{K}$ signal may be used to constrain
$m_{G_0}$.
 \item
Lattice QCD suggests that the primitive scalar glueball $G_0$  lies
at or above 1500 MeV, hence above the $I=1$ $Q\bar{Q}$ state
$a_0(1450)$ and the (presumed) associated $n\bar{n}$ $f_0(1370)$.
Hence $E_{G_0}-E_{n\bar{n}} >0$ in the numerator of $\omega$, eqn. 6.
\item
The allowed range for $\omega ($ eqn. \ref{range})
enables predictions of the mass of the $\Psi_s$ state.
Using eqn. \ref{omega} and $m(G) = 1509 \pm 10 $ MeV,
$m(\Psi_n) =
1360 \pm 40$ MeV, which dominates the error, we find \cite{cafe}
\begin{equation}
1580 [m(\Psi_n) = 1400] < m(\Psi_s) < 1890 [m(\Psi_n) = 1320]
\ {\rm MeV}
\end{equation}
This is consistent with naive mass estimates whereby the
 $\Delta m = m_{s\bar{s}} - m_{n\bar{n}} \approx 200-300$ MeV
\end{enumerate}
If this were the whole story, the mixing eqn. \ref{3states} would suppress
not just $K\bar{K}$ but also $\eta\eta$. However, this is where the
effect of the perturbation $V_2$ comes into prominence: At $O(V_2)$
the glueball decays directly into pairs of glueballs or $(gg)$
continuum and thereby into mesons
whose Fock states have strong overlap with $gg$.
 To the
extent that there is significant $gg$ coupling to $\eta,\eta^\prime$
or to $\sigma \equiv (\pi\pi)_s$ (e.g. $\psi' \rightarrow \psi gg
\rightarrow \psi \eta$
and $\psi (\pi\pi)_s$ have large intrinsic couplings
notwithstanding the fact that they are superficially OZI violating) one
may anticipate $\eta\eta$ and $\eta\eta^\prime$
 in the two-body decays of scalar glueballs. Note that the perturbation $V_2$
triggering $G_0 \rightarrow G_0G_0$, $ G_0 (gg)$ or $(gg)(gg)$
will not directly feed exclusive two-body flavoured states such as
$\pi\pi$ nor $K\bar{K}$ since gluons are isoscalar.
However, from the above $\psi$ phenomenology, one may anticipate
$(gg)(gg) \rightarrow (\pi\pi)_s(\pi\pi)_s$ in the decay of $G(0^{++})$
and analogously
for $0^{-+}$ glueballs one may anticipate $ \eta\sigma$ or
$\eta'\sigma$ decays.

The manifestation of this mechanism in final states involving
the $\eta$ or $\eta'$ mesons depends on the unknown overlaps
such as $\langle gg|V|q\bar{q} \rangle$ in the pseudoscalars. Chiral
symmetry suggests that the coupling of glue to $\eta$ or $\eta'$
vanishes in the limit $m_q \rightarrow 0$ and hence occurs dominantly
through their $s\bar{s}$ component, thereby favouring the $\eta'$:

\begin{equation}
\frac{\langle gg|V|\eta' \rangle}{\langle gg|V|\eta \rangle}
 = \frac{\langle s\bar{s}|\eta' \rangle}{\langle s\bar{s}|\eta \rangle}
\sim
- \frac{4}{3}.
\label{chir}
\end{equation}

This appears to be consistent with the decays of $f_0(1500)$ as we now
show. Combining the $\eta\eta$ amplitudes from both the $Q\bar{Q}$
and $G_0$ components we have, in the approximation that $\eta$ and
$\eta'$ are approximately $50:50$ mixtures of $n\bar{n}$ and
$s\bar{s}$ \footnote{The actual numbers vary slightly when
a mixing angle of -17.3$^o$ \cite{PS} is used, see ref. \cite{cafe}}
\begin{equation}
\frac{\langle\eta\eta |V|G\rangle}{\langle\pi\pi |V|G\rangle} =
\frac{\langle\eta\eta |V|G_0 \rangle}{N_G\langle\pi\pi |V|G\rangle}
 + (\frac{1+\omega }{2})
= \pm(0.90 \pm 0.20)
\label{a1}
\end{equation}
from $R_1$ (eqn. \ref{R1}) and
\begin{equation}
\frac{\langle\eta\eta^\prime  |V|G\rangle}{\langle\pi\pi |V|G\rangle}=
\frac{\langle\eta\eta^\prime  |V|G_0\rangle}{N_G\langle\pi\pi
|V|G\rangle}
 + (\frac{1-\omega}{2}) = \pm (0.53 \pm 0.11)
\label{a2}
\end{equation}
from $R_2$ (eqn. \ref{R2}).
Hence using the range eqn. \ref{range} for $\omega $ one finds
with the $+$ sign in eqn. \ref{a1} and the $-$ sign in eqn. \ref{a2}

\begin{equation}
- 1.8 < \frac{\langle \eta \eta'|V|G_0 \rangle}{\langle \eta\eta | V|G_0
\rangle}
 < -1.5.
\end{equation}
A small breaking of chiral symmetry is consistent with this.

An upper limit for $\xi^2$ can be obtained by comparing the
ratio of amplitudes (eqn. 13) for $f_0(1500)$ decay to that for
$f_0(1370)$. We find \cite{cafe}: $\xi^2 < $ 0.2.
The magnitudes of the partial widths of $1500/1370$ also fit with the
$G-Q\bar{Q}$ mixing scenario. The above analysis shows that the decay
amplitudes from the $Q\bar{Q}$ component of
are all at $0(\xi)$ and so we expect
\begin{equation}
\frac{\gamma^2 (f(1500)\rightarrow\pi\pi)}{\gamma^2
(f(1370)\rightarrow\pi\pi)} \simeq
O(2\xi^2).
\end{equation}
Empirically this ratio is smaller than 0.4
and hence we again find $\xi^2 \leq 0.2$.

This hypothesis implies that there are small $G_0$ admixtures in
$\Psi_{n,s}$:

\begin{eqnarray}
\sqrt{1+\omega^2 \xi^2} |\Psi_{s} \rangle =
|s\bar{s} \rangle - \xi  \omega |G_0 \rangle
\nonumber\\
\sqrt{1+2\xi^2} |\Psi_{n} \rangle =
|n\bar{n} \rangle - \xi \sqrt{2} |G_0 \rangle
\nonumber\\
\label{3satates}
\end{eqnarray}
which in turn implies that the sum of the partial widths of the two
states is
\begin{equation}
\Gamma(\Psi_n ) + \Gamma(G) \simeq \Gamma (n\bar{n}),
\end{equation}
in accord with quark model estimates \cite{Godfrey,Barnes}.

The smallness of $|\xi|$ then also implies that $\Psi_s$ decays
essentially like an $s\bar{s}$ state
with the branching ratios to $K\bar{K}$ and $\eta\eta'$ dominating over
$\eta\eta$ and with $\pi\pi$ much suppressed. If $^3P_0$ quark pair creation
is important in the decay dynamics,
the values of the branching ratios and total width may be strongly mass
dependent \cite{cafe,kokoski87,Barnes}

The quantitative predictions of our analysis depend on the apparent
suppression of $f_0(1500)$ decay to $\KKbar$. Thus detailed study of
$p\bar{p}\rightarrow\pi K\bar{K}$ can be seminal (i) in confirming
the $\KKbar$ suppression, (ii) in confirming the $K^*_0$ (1430)
$\rightarrow
K\pi$ and $a_0(1450)\rightarrow\eta\pi$ and $K\bar{K}$,
(iii) in quantifying the signal for $f_0$ (1370) and $f_0$ (1500)
and (iv) in isolating the predicted $\ssbar$ member of the nonet.
Furthermore, study of radiative decays $f_0(1500) \rightarrow
\gamma + \omega (\phi,\rho)$
may probe the flavour content in the $Q\bar{Q}$ mixing.
To the extent that $\omega \equiv n\bar{n}$ and $\phi \equiv
s\bar{s}$, the amplitude ratios will be
\begin{equation}
f_0(1500) \rightarrow \gamma\phi : \gamma \omega: \gamma \rho =
-\omega  : 1 : 3.
\end{equation}

It seems clear that new dynamics, beyond simple $Q\bar{Q}$, is
operating at the $f_0(1500)$. The simplest explanation within strong
QCD is that a glueball excitation is seeding the phenomena, in particular
the clarity of the signals in channels that are believed favourable to
glue. It is also tantalising that the signal appears particularly sharp in
$p\bar{p} \rightarrow \eta\eta\eta$ \cite{CBunpub} which might be
consistent with direct production of a $0^{-+}$ glueball above 2GeV (in
line with lattice predictions) decaying via the $V_2$ process into
$G_0G_0$ with consequent affinity for $f_0(1500) + \eta$. An excitation
curve for this process could be interesting.

We thank T. Barnes, M. Benayoun, K.Bowler,
D. Bugg, Y. Dokshitzer, G. Gounaris, A. Grigorian, R.Kenway,  E. Klempt,
H.J Lipkin, J. Paton, S. Spanier, M. Teper, D. Wyler and B. Zou for helpful
discussions.

\pagebreak
\section*{Figure Captions}

\begin{figure}[h]
\vspace{1mm}
\caption[]{The effect of perturbation $V_1$ causes $Q\bar{Q}$ decay (a)
or $G_0$ mixing with $\qqbar$ (b). The effect of $V_2$ on $G_0$ causes
decay to two glueballs (c).}
\label{Rivsx}
\end{figure}

\begin{figure}[h]
\vspace{1mm}
\caption[]{Invariant couplings $\gamma^2$ to two pseudoscalars
for an isoscalar $0^{++} \qqbar$ meson as a function of nonet
mixing angle for a pseudoscalar mixing angle of $-17.3^{\circ}$
(up to a common arbitrary normalization constant).
Full curve: $\eta\eta$; dashed curve:
$\eta\eta'$; dotted curve: $\pi\pi$; dashed-dotted curve:
$\KKbar$. Ideal mixing occurs at 35.3$^{\circ}$ ($\ssbar$) or at
125.3$^{\circ}$ ($\nnbar$).}
\label{Glas}
\end{figure}

\begin{figure}[h]
\vspace{1mm}
\caption[]{Gluonium-$Q\bar{Q}$ mixing
involving the energy denominator $E_G-E_{Q\bar{Q}}$}
\label{3graphs}
\end{figure}

\end{document}